\documentstyle[11pt]{article}
\begin{document} \date{}
\newcommand{\ba}{\begin{array}}
\newcommand{\ea}{\end{array}}
\newcommand{\bea}{\begin{eqnarray}}
\newcommand{\eea}{\end{eqnarray}}
\newcommand{\sta}{\stackrel}
\newcommand{\ovl}{\overline}
\newcommand{\eee}{\mbox{e}}
\newcommand{\triex}{\mbox{\hspace{3ex}}}
\newcommand{\lla}{\longleftarrow}
\newcommand{\lra}{\longrightarrow}
\newcommand{\edoc}{\end{document}}
\newcommand{\Pfaff}{\mbox{Pfaff}}
\newcommand{\ch}{\mbox{ch}\,}
\newcommand{\sh}{\mbox{sh}\,}
\newcommand{\th}{\mbox{th}\,}
\newcommand{\pal}{\partial}
\newcommand{\spsigma}{\ba[t]{c} \mbox{Sp} \vspace{-1ex} \cr
\mbox{$\scriptstyle{(\sigma)}$} \ea }
\newcommand{\spsigmamn}{\ba[t]{c} \mbox{Sp} \vspace{-1ex} \cr
\mbox{$\scriptstyle{(\sigma_{mn})}$} \ea }
\newcommand{\spursigmamn}{\ba[t]{c} \mbox{Sp} \vspace{-1ex} \cr
\mbox{$\scriptstyle{(\sigma_{mn})}$} \ea }
\newcommand{\spura}{\ba[t]{c} \mbox{Sp} \vspace{-1ex} \cr
\mbox{$\scriptstyle{(a)}$} \ea }
\newcommand{\spurasigma}{\ba[t]{c} \mbox{Sp} \vspace{-1ex} \cr
\mbox{$\scriptstyle{(\sigma\,|\,a)}$} \ea }
\newcommand{\spuramn}{\ba[t]{c} \mbox{Sp} \vspace{-1ex} \cr
\mbox{$\scriptstyle{(a_{mn})}$} \ea }
\newcommand{\spurbmn}{\ba[t]{c} \mbox{Sp} \vspace{-1ex} \cr
\mbox{$\scriptstyle{(b_{mn})}$} \ea }
%
%
\title{\LARGE\bf Fermionic Path Integrals and Two-Dimensional Ising
Model with Quenched Site Disorder }
\author{V.N. Plechko \\[3mm]
\em  Bogoliubov Laboratory of Theoretical Physics,\\
\em  Joint Institute for Nuclear Research, JINR-Dubna,\\
\em  141980 Dubna, Moscow Region, Russia \\
     E-mail: plechko@thsun1.jinr.ru}
\maketitle

\begin{abstract}  \noindent The notion of the integral over the
anticommuting Grassmann variables is applied to analyze the fermionic
structure of the 2D Ising model with quenched site dilution. In the
$N$-replica scheme, the model is explicitly reformulated as a theory of
interacting fermions on a lattice. For weak dilution, the continuum-limit
approximation implies the log-log singularity in the specific heat near
$T_c$. \footnote{ \  Published in ``Path Integrals from peV to TeV:
50  Years after Feynman's Paper'',  Proceedings of the 6th International
Conference on Path Integrals from peV to TeV, August 25--29, 1998,
Florence, Italy.  Edited by  R. Casalbuoni, R. Giachetti, V. Tognetti, R.
Vaia and P. Verrucchi (World Scientific, Singapore, 1999) p.p.~137--141.
HEP-TH/9906107 } \end{abstract} \vspace{3ex}

%
%
%
\section{Introduction}

The site dilution provides, probably, the simplest way to realize quenched
disorder in real magnetic materials. In this case, some amount of the
magnetic atoms in a sample, chosen at random, are replaced by nonmagnetic
impurity atoms. Another sort of disorder is bond dilution, in which case
some of the lattice bonds are assumed to be broken.  The two-dimensional
Ising model (2DIM) is a natural object to analyze the effects of disorder
on ferromagnetic phase transition since the exact analytic solution is
known for this model in the pure case \cite{ons44}. The disordered 2D
Ising model already has been analyzed during the last decades both by
theoretical tools and in the precise Monte-Carlo experiments
\cite{stinch83,dd83,bns94,vid95,sst94}. The theoretical studies were merely
concerned with the case of bond dilution, making use of the fermionic
interpretation of the problem \cite{dd83,bns94,vid95}. In this report, we
apply a new anticommuting path integral technique \cite{ber66,ple85,ple88}
to clarify the fermionic structure of the 2DIM with quenched site
dilution \cite{ple98}. At the first stage, the partition function with
fixed dilution is transformed into a fermionic Gaussian integral. The
averaging over the disorder within the $N$-replica scheme then results a
lattice fermionic theory with interaction.  Unexpectedly, the interaction
in the exact lattice theory appears to be of order $2N$ in fermions, where
$N \to 0$ is the number of replicas \cite{ple98}. An effective
four-fermion interaction arises, however, in a continuum-limit field theory
responsible for the critical behaviour near $T_c\,$ for weak dilution due
to the interplay of the short-wave and long-wave lattice fermionic modes as
we pass to the continuum limit. In particular, this implies the
double-logarithmic singularity in the specific heat near $T_c$ for weak
site dilution. The question about the effects of strong and moderated
dilution, however, still remains to be an important open problem in the 2D
Ising model, this holds true for both site and bond dilution.

%
%
\section{ The model}

Let us start with a fixed distribution of the nonmagnetic sites over a
lattice (fixed site dilution). The Ising spins, $\sigma_{mn}=\pm1$, are
located at lattice sites, $mn$, with $m,n= 1,2,...,L\,$ running in the
horizontal and vertical directions, respectively. Here $L$ is the lattice
length, $L \to \infty$. To introduce site dilution, we accompany each Ising
spin by a random variable $y_{mn}=0,1$. The hamiltonian is \cite{ple98}:
\bea
-\beta H\,\{y\,|\,\sigma\} =\sum\limits_{mn}^{}
\left[\,b_{1}\,y_{mn}y_{m+1n}\sigma_{mn}\sigma_{m+1n}
+ b_{2}\,y_{mn}y_{mn+1}\sigma_{mn}\sigma_{mn+1}\right]\,,
\label{ham1}
\eea
where $b_{1,2} = \beta J_{1,2}$ are the bond coupling parameters, $J_{1,2}$
are the exchange energies, $\beta=1/kT$ is the inverse temperature. We
assume ferromagnetic case: $b_{1,2}\!>\!0$. Noting the identity for a
typical bond weight: $\exp\,(b\,yy' \sigma\sigma') =\cosh(b\,yy') +yy'
\sigma\sigma'\sinh(b\,yy')$, which readily follows from $\sigma\sigma'
=\pm1$ and $yy'=0,1$, the partition function can be written in the form:
$Z\,\{\,y\,\}=R\,\{\,y\,\} \,Q\, \{\,y\,\}$, where $R\{y\}$ is a
nonsingular spin-independent prefactor, to be ignored in what follows, and
$Q\{y\}$ is the reduced partition function \cite{ple98}:
\bea
Q\{y\}=\spsigma\!\prod\limits_{\,mn}\;
(1+t_{1}\,y_{mn}y_{m+1n}\sigma_{mn}\sigma_{m+1n})
(1+t_{2}\,y_{mn}y_{mn+1}\sigma_{mn}\sigma_{mn+1})\,,\mbox{ \ }
\label{qss1}
\eea
where $t_{1,2} =\tanh\,b_{1,2}\,$ and a properly normalized spin
averaging is assumed. Since we are interesting in quenched disorder, we
have to average the free energy rather than the partition function itself.
The standard device to avoid the averaging of the logarithm is the replica
trick:
\bea
\ovl{\stackrel{}{-\beta f_{\,Q\,}\{\,y\,\}}}
= \ovl{\stackrel{}{\ln\,Q\{\,y\,\}}}
= \lim_{N \to 0}\,\frac{1}{N}\,
\ovl{\stackrel{}{(Q^N\{y\}-1)}}\,.
\label{repl1}
\eea
In this scheme, we take $N$ identical copies of the original partition
function and average $Q^{\,N}\{y\}\,$. The formal limit $N\to0$ to be
performed at final stages. In what follows, we assume the simplest
distribution of the impurities in the definition of the averaging:
\bea
W\,(y_{mn}) = p\,\delta\,(1 - y_{mn}) + (1 - p)\,\delta\,(y_{mn})\,,
\label{aver1}
\eea
where $\delta(\ )$ are the correspondent Kronecker's symbols, $p$ is the
probability that any given site, chosen at random, is occupied by the
normal Ising spin, while $1-p$ is the probabilty that the given site is
dilute. \vspace{2pt}

%
%
\section{Fermionic interpretation}

The partition function with fixed
disorder (\ref{qss1}) can be transformed into a Gaussian fermionic
integral \footnote{       \ The rules of the integration over the purely
anticommuting (Grassmann) variables were first introduced by F.A.
Berezin \cite{ber66}. In a close analogy with the bosonic case, a Gaussian
fermionic integral of any kind can be expressed in terms of the determinant
of the associated matrix \cite{ber66}. For a short comment about the
Gaussian fermionic integrals also see \cite{ple96,hp97}.} following the
method of the mirror-ordered factorization for the density
matrix.\,\footnote{ \ The starting point of the method is the fermionic
factorization of the local spin-variable bond Boltzmann weights,
$\left<\sigma\,|\, \sigma'\right>$.  Introducing intermediate fermionic
variables, we write $\left<\sigma\,|\, a\right> \left<a\,|\,\sigma'
\right>$. After elimination of the spin variables, it is then possible to
express the partition function in terms of the purely fermionic factors,
$\left<a\,|\, a' \right>$. In fact, the mixed factors like $\left<\sigma\,
|\,a\right>$, arising by inserting the fermions, are neither commuting nor
anticommuting with each other. So, one have to invent a special ordering
for these factors, in their global products, in order the elimination of
spin variables be really possible \cite{ple85,ple88}. In particular, the
ordering problem is an obstacle to deal with the 3D Ising model. In two
dimensions, however, the method is simple \cite{ple85,ple88}. Following
this way, we come to (\ref{qab1}) \cite{ple98}.}\,\,\cite{ple85,ple88}.
This results the integral \cite{ple98}:
\bea
&& Q\{y\} = \int
\prod\limits_{mn}^{}db_{mn}^{\,*}db_{mn}^{}
da_{mn}^{\,*}da_{mn}\exp\sum\limits_{mn}^{}
\Big\{ a_{mn}^{}a_{mn}^{\,*}+b_{mn}^{}b_{mn}^{\,*}\,+
\label{qab1} \\ \nonumber
&& +\, y_{mn}^{2}\Big[\,a_{mn}^{}b_{mn}^{}+ \,t_{1}t_{2}\,
a_{m-1n}^{\,*}b_{mn-1}^{\,*}+\,(t_{1}a_{m-1n}^{\,*}
+\, t_{2}b_{mn-1}^{\,*})(a_{mn}^{}+b_{mn}^{})\,\Big]\Big\}\,,
\eea
where $a_{mn}, a_{mn}^{\,*}, b_{mn}, b_{mn}^{\,*}$ are Grassmann variables.
In turn, integrating out a part of fermionic variables from (\ref{qab1}),
we obtain the reduced integral \cite{ple98}:
\bea
&& Q\,\{y\}= \int \prod\limits_{mn}^{} d\bar{c}_{mn}^{}
dc_{mn}^{}\,y_{mn}^{\,2}\,\exp\sum\limits_{mn}^{}\,
\Big[\,y_{mn}^{-2}\,c_{mn}^{}\bar{c}_{mn}^{}\,+
\cr
&& +\,(c_{mn}^{}+\bar{c}_{mn}^{})\,(t_{1}c_{m-1n}^{} -
t_{2}\bar{c}_{mn-1}^{})- y_{mn}^{\,2}\,t_{1}t_{2}\,
c_{m-1n}^{}\bar{c}_{mn-1}^{}\,\Big]\,,
\label{qcc1}
\eea
where $c_{mn},\bar{c}_{mn}$ are Grassmann variables, $y_{mn}^{\,2}\,
\exp\,(\,y_{mn}^{-2}c_{mn}\bar{c}_{mn})= y_{mn}^{\,2}+c_{mn}\bar{c}_{mn}$.
The integrals (\ref{qab1}) and (\ref{qcc1}) are completely equivalent to
each other and to (\ref{qss1}). The advantage of the reduced
representation like (\ref{qcc1}) is that the corresponding action, in the
pure case, explicitly illuminates the Majorana-Dirac structures of 2DIM
already at the lattice level \cite{ple98}.

The averaging of (\ref{qcc1}) over the disorder within the $N$-replica
scheme, see (\ref{repl1}) and (\ref{aver1}), results the theory with
interaction of the following kind \cite{ple98}:
\bea
\ovl{\stackrel{}{Q^N\{y\}}} =
\int \prod\limits_{mn}^{}\prod\limits_{\alpha=1}^{N}
d\bar{c}_{mn}^{\;(\alpha)}dc_{mn}^{\,(\alpha)}\,\prod\limits_{mn}^{}
\Big[\,p\,\prod\limits_{\alpha=1}^{N}
\eee^{\,S_{mn}^{\,(\alpha)}} + (1-p)\,\prod\limits_{\alpha=1}^{N}
c_{mn}^{\,(\alpha)}\bar{c}_{mn}^{\;(\alpha)}\;\Big]\;\;&&
\label{qnn1} \\ \nonumber
=p^{L^2}\!\!\int\prod\limits_{mn}\prod\limits_{\alpha=1}^{N}
d\bar{c}_{mn}^{\;(\alpha)}dc_{mn}^{\,(\alpha)}\,\exp\,
\sum\limits_{mn}^{}\Big[\sum\limits_{\alpha=1}^{N}
S_{mn}^{\,(\alpha)}+ \frac{1-p}{p}\prod\limits_{\alpha=1}^{N}
c_{mn}^{\,(\alpha)}\bar{c}_{mn}^{\;(\alpha)}\eee^{\,
\Delta_{mn}^{\,(\alpha)}}\Big]\,,\,&&
\eea
where $S_{mn}^{\,(\alpha)}$ is the replicated Gaussian action from
(\ref{qcc1}) for the pure case, with $y_{mn} \equiv 1$ at all sites, and
$\Delta_{mn}^{\,(\alpha )} =\,-\,S_{mn}^{\,(\alpha)}$. Taking into account
the nilpotent property of fermions in prefactor before the exponential,
another possible choice is: $\Delta_{mn }^{\, (\alpha)} =t_{1}t_{2}\,
c_{m-1n}^{\,(\alpha)} \bar{c}_{mn-1}^{\,(\alpha)}$ \cite{ple98}.
The continuum-limit field theory for weak dilution that follows from
(\ref{qnn1}) is commented in the next section. A re\-la\-tively simple
form of interaction in (\ref{qnn1}) also provides grounds to try the cases
of strong and moderated dilution, starting directly with the exact lattice
integral (\ref{qnn1}).  The analysis of this kind, however, have not yet
been performed.\vspace{2pt}

%
%
\section{The Gross-Neveu model ($N \to 0$)}

To extract the effective continuum-limit field theory from (\ref{qnn1}),
we have to distinguish explicitly the higher- and low-momentum lattice
fermionic modes (Fourier-harmonics) in the exact lattice action
(\ref{qnn1}). Integrating out the higher-momentum modes in the first order
of perturbation theory (weak dilution), we obtain an effective action for
the low-momentum fields (large distances). The effective action appears,
finally, in the form of the $N$-colored Gross-Neveu model ($N\to0$)
\cite{ple98}:
\bea
&& S_{\rm\,G-N} =
\int d^2x \Big\{\sum\limits_{\alpha=1}^{N}\Big[\,
\ovl{m}_N\,\psi_{1}^{(\alpha)}\psi_{2}^{(\alpha)}+\,\frac{1}{2}\,
\psi_{1}^{(\alpha)}\,(\pal_1+i\,\pal_2)\,\psi_{1}^{(\alpha)}
\label{gross1}
\\ \nonumber
&& +\, \frac{1}{2}\,\psi_{2}^{(\alpha)}\,(-\pal_1+i\,\pal_2)\,
\psi_{2}^{(\alpha)}\Big] + g_{\mathstrut N} \Big[\sum\limits_{\alpha=1}^{N}
\psi_{1}^{(\alpha)}\psi_{2}^{(\alpha)}\Big]^{\,2}\,\Big\}\,,
\\ \nonumber
&& \ovl{m}_{\mathstrut N}\,
= {1-t_1-t_2-t_1\,t_2 \over \sqrt{2(t_1t_2)_c}}
+ \left<A\right>^N\,\frac{1-p}{p}\,\frac{\left<B\right>}{\left<A\right>}\,
\frac{1}{\sqrt{2\,(t_1t_2)_c}}\,,
\\ \nonumber
&& g_{\mathstrut N}\,=\,\left<A\right>^N\,
\frac{1-p}{p}\,\frac{\left<B\right>^2}{\left<A\right>^2}\,
\frac{1}{4\,(t_1t_2)_c}\,,
\eea
where $\psi_1,\psi_2$ are the anticommuting Majorana components,
$\ovl{m}_N$ and $g_N$ are the effective mass and charge, respectively.
Here $\left<A\right>$ and $\left<B\right>$ are some lattice fermionic
averages (short distances) explicitly calculated in \cite{ple98}. The
Gaussian part in (\ref{gross1}) is the replicated Majorana action,
corresponding to the pure case, with the mass term modified by disorder.
The $N=0$ G-N model similar to (\ref{gross1}) already has been analyzed
intensively by DD-SSL as an effective theory near $T_c$ for weak bond
dilution \cite{dd83,bns94,vid95,dpp95}. The predictions are the
double-logarithmic singularity in the specific heat and the logarithmic
corrections in other thermodynamic functions. For more details see
\cite{bns94,vid95,aarao97,roder98}. The present analysis thus confirms the
idea of the universality for small amount of impurities in the 2D Ising
ferromagnets, for weak dilution.  The situation is less evident, however,
for strong and moderated site and/or
bond dilution \cite{sst94,aarao97,roder98}. \vspace{2pt}

\section{ Added note}

Let $f\,(c_1,c_2,...,c_N)$ be any function of Grassmann variables $c_1,c_2,
..,c_N$. From the basic rules of fermionic integration, it follows:
\bea
&& \int dc_N...dc_2dc_1\,f\,(\lambda_1c_1,\lambda_2c_2,...,\lambda_Nc_N)
\cr
&& =\lambda_1\lambda_2...\lambda_N \!\int\!
dc_N...dc_2dc_1\,f\,(c_1,c_2,...,c_N)\,.
\label{iden1}
\eea
Infinite series of related identities also follows by differentiating
(\ref{iden1}) with respect to the parameters $\lambda_1, \lambda_2, ...,
\lambda_N$. In particular, applying these ideas to the integral
(\ref{qnn1}), we obtain the identity:
\bea
\frac{1}{L^{\,2}}\sum\limits_{mn}^{}
\Big<\,S_{mn}^{\,(\alpha)} + \mbox{$\frac{1-p}{p}$}\;
\eee^{\,\Delta_{mn}^{(\alpha)}}\, \mbox{$\prod\limits_{\beta=1}^{N}$}
\,c_{mn}^{\,(\beta)}\bar{c}_{mn}^{\;(\beta)}\,
\eee^{\,\Delta_{mn}^{(\beta)}}\Big> =1\,,
\label{sumrul1}
\eea
where the averaging is assumed to be taken with respect to the non-Gaussian
fermionic measure from (\ref{qnn1}). This identity can be used to check the
consistency of approximations of any kind when dealing with lattice theory
(\ref{qnn1}). In a sense, the equation (\ref{sumrul1}) can be viewed as the
analog of $\left<p^2/2m \right>= \theta/2$, $\theta=kT$, for the kinetic
energy of a particle in classic Maxwell gas (one degree of freedom). In any
case, the later identity readily follows by exactly the same
method.\vspace{-5pt}

%
%
\section*{Acknowledgments:}

The discussions with B.N. Shalaev are gratefully acknowledged. I would like
to thank the organizers of PI-98 for their kind hospitality during the
conference. This work was supported in part by RFFI grant 98-02-10897.

\mbox{}


\vspace{-7pt}

\begin{thebibliography}{45}

\bibitem{ons44}  L. Onsager, Phys. Rev. 65 (1944) 117.

\bibitem{stinch83}
R.B. Stinchcombe, in Phase Transitions and Critical Phenomena, vol.~7,
ed. by C. Domb and J.L. Lebowitz (Academic Press, London, 1983).

\bibitem{dd83}
V.S. Dotsenko and V.S. Dotsenko, Adv. Phys. 32 (1983) 129.

\bibitem{bns94} B.N. Shalaev, Phys. Rep. 237 (1994) 129.

\bibitem{vid95}  Vik. S. Dotsenko, Sov. Phys. Usp. 38 (1995) 310.

\bibitem{sst94}
W. Selke, L.N. Shchur and A.L. Talapov,  Monte Carlo Simulations of Dilute
Ising Models, in  Annual Reviews of Computational Physics I, ed. by  D.
Stauffer (World Scientific, Singapore, 1994), p. 17.

\bibitem{ber66}
F.A. Berezin, The Method of Second Quantization (Academic Press,
New York, 1966).

\bibitem{ple85}
V.N. Plechko, Sov. Phys. Doklady, 30 (1985) 271.

\bibitem{ple88}
V.N. Plechko, Physica A 152 (1988) 51.

\bibitem{ple98}
V.N. Plechko, Phys. Lett. A 239 (1998) 289; (E) 245 (1998) 563.

\bibitem{ple96}
V.N. Plechko, A talk given at the V~Int.\ Conf.\ on Path Integrals from
meV to MeV, Dubna, Russia, May 27--31, 1996. \ hep-th/9609044.

\bibitem{hp97}
R. Hayn and V.N. Plechko, A talk given at the VIII~Int.\ Conf.\ on Symmetry
Methods in Physics,  Dubna, Russia, July 28 -- August 2, 1997. \
cond-mat/9711156.

\bibitem{dpp95}
Vl. Dotsenko, M. Picco and P. Pujol. Phys. Lett. B 347 (1995) 113.

\bibitem{aarao97}
F.D.A. Aarao Reis, S.L.A. de Queiroz and R.R. dos Santos, Phys. Rev.
B 56 (1997) 6013.

\bibitem{roder98}
A. Roder, J. Adler and W. Janke, Phys. Rev. Lett. 80 (1998) 4697. \\
J.\,-K. Kim, cond-mat/9808280.

\end{thebibliography}
\end{document}